\documentclass[osajnl,twocolumn,showpacs,superscriptaddress,10pt]{revtex4-1} 
\usepackage{amsmath,amssymb,graphicx}
\usepackage{color}
\newcommand{\tE}{\tilde{E}}
\newcommand{\trho}{\tilde{\rho}}
\newcommand{\EE}{\mbox{\boldmath $E$}}

\newcommand{\tgamma}{\tilde{\gamma}}

\begin{document}

\renewcommand{\deg}{^{\circ}}
\newcommand{\micron}{\mu\mbox{m}}

\newcommand{\WASEDA}{\affiliation{Department of Applied Physics,
    School of Advanced Science and Engineering, Waseda University,
    3-4-1 Okubo, Shinjuku-ku, Tokyo 169-8555, Japan}}
\newcommand{\KANAZAWA}{\affiliation{Faculty of Mechanical Engineering,
    Institute of Science and Engineering, Kanazawa University,
    Kakuma-machi, Kanazawa, Ishikawa 920-1192, Japan}}
\title{Universal single-mode lasing in fully chaotic two-dimensional microcavity lasers under continuous-wave operation with large pumping power}

\author{Takahisa Harayama}\WASEDA
\author{Satoshi Sunada}\KANAZAWA
\author{Susumu Shinohara}\WASEDA

\begin{abstract}
For a fully chaotic two-dimensional (2D) microcavity laser, we present a theory that guarantees both the existence of a stable single-mode lasing state and the nonexistence of a stable multimode lasing state, under the assumptions that the cavity size is much larger than the wavelength and the external pumping power is sufficiently large. It is theoretically shown that these universal spectral characteristics arise from the synergistic effect of two different kinds of nonlinearities: deformation of the cavity shape and mode interaction due to a lasing medium. Our theory is based on the linear stability analysis of stationary states for the Maxwell-Bloch equations and accounts for single-mode lasing phenomena observed in real and numerical experiments of fully chaotic 2D microcavity lasers.

\vspace{0.3cm}
\noindent
{\it OCIS codes}:
(140.3945) Micorcavities;
(140.3410) Laser resonators;
(270.3430) Laser theory;
(000.1600) Classical and quantum physics.

\vskip 5mm
\noindent
\href{https://doi.org/10.1364/PRJ.5.000B39}{doi:10.1364/PRJ.5.000B39}
\vskip 3mm
\noindent
\copyright~2017~Optical Society of America. Users may use, reuse, and
build upon the article, or use the article for text or data mining, so
long as such uses are for non-commercial purposes and appropriate
attribution is maintained. All other rights are reserved.

\end{abstract}

\maketitle 

\section{Introduction}
Universality is a key concept in quantum chaos study. 
We can find a common feature in a quantum system if the corresponding
classical system exhibits fully chaotic dynamics \cite{StoeckmannBook,
  Haake, NakamuraHarayama}.
The most representative example is the Bohigas-Giannoni-Schmit
conjecture concerning universal spectral fluctuations of quantized
fully chaotic systems \cite{BohigasCasatiBerry, MullerHeusler}.
Universality in quantum chaos has also been observed for electron
transport in mesoscopic devices \cite{Jalabert}. 

Because of an analogy between the classical-quantum and the ray-wave
correspondence, quantum chaos theory can be directly applied to
``wave-chaotic'' systems, which exhibit chaotic dynamics in the
ray-optic limit \cite{StoneNoeckeletc}.
A representative example is a two-dimensional (2D) microcavity laser,
whose resonant modes can be viewed as those of quantum billiards
\cite{JanCao, HarayamaShinohara}.
By a quantum-chaos approach, the emission patterns of 2D microcavity
lasers with various cavity shapes have been successfully explained and
predicted \cite{StoneNoeckeletc, JanCao, HarayamaShinohara}.
However, there is another important aspect of 2D microcavity lasers
that cannot be elucidated only by quantum chaos theory, that is,
nonlinear interactions among resonant modes due to a laser gain
medium.
From the viewpoint of universality, it is of interest to uncover how
this additional nonlinear effect manifests itself depending on the
chaoticity or integrability of underlying ray dynamics inside a
cavity.

A recent experimental study of semiconductor 2D microcavity 
lasers has demonstrated that single-mode lasing is achieved with 
a stadium-shaped (i.e., fully chaotic) cavity, while multimode lasing 
with an elliptic (i.e., nonchaotic) cavity \cite{Sunada1, Sunada2}.
This drastic difference was attributed to the difference of 
spatial modal patterns between the stadium and elliptic cavities. 
It was numerically shown that for the stadium cavity, an arbitrary 
low-loss modal pair has a significant spatial overlap, while for  
the elliptic cavity, there exist low-loss modal pairs whose spatial 
overlaps are small \cite{Sunada2}. This means that any lasing mode 
of the stadium cavity tends to strongly interact with the other modes, 
while multiple lasing modes can coexist for the elliptic cavity 
because of small interactions among them.

It is important that all of the various fully chaotic stadium-type
cavities of different aspect ratios and sizes studied in the
experiments of Ref.\cite{Sunada2} have shown single-mode lasing while
all of various integrable elliptic cavities multimode lasing.
Therefore, it was conjectured that single-mode lasing is universal for
fully chaotic cavity lasers \cite{Sunada2}.  In order to ascertain
this universality, it is important to further examine it
experimentally and numerically for various 2D cavities.
Such studies are important because they would add evidence of the universality. 
However, these pieces of evidence cannot directly reveal 
the insight of the universality. 
In past studies of universalities not only in quantum chaos
but also in second-order phase transitions and critical phenomena,
establishment of a theory that verifies a universality was the
most difficult, challenging and important task 
\cite{MullerHeusler, Fisher}.

For a theoretical verification of the conjecture of universal
single-mode lasing, it is necessary to evaluate the stability of a
stationary-state solution of a full nonlinear model such as the
Maxwell-Bloch equations \cite{Harayama2003, Harayama2005,
  Tureci2006,Tureci2007}.  Stability analysis for one-dimensional
lasers in a low pumping regime just above the lasing threshold has
been established by Lamb \cite{Lamb, L2, Sargent}.  It also has been
applied to 2D microcavity lasers and explained the spontaneous
symmetry breaking of a lasing pattern \cite{HarayamaAsymmetricPRL,
  Harayama2005}.
However, Lamb's perturbation theory becomes invalid for a high pumping
regime where the universal single-mode lasing can be observed because
his perturbation theory expresses the population inversion by a power
series expansion of lasing modes.

In this paper, we introduce a different expansion method for the
population inversion in the Maxwell-Bloch equations that is applicable
to a high pumping regime.  Furthermore, we explicitly derive the
stability matrix for stationary-state solutions, which describes the
interactions among a huge number of lasing modes. The matrix elements
turn out to be greatly simplified under the assumptions that the
cavity size is much larger than the wavelength and the external
pumping power is sufficiently large.  Moreover, the eigenvalues of
this matrix can be analytically evaluated by applying a theorem in
linear algebra. This enables us to theoretically show that for a fully
chaotic 2D microcavity laser, at least one single-mode lasing state is
stable, while all multimode lasing states are unstable.
This result provides a theoretical ground for 
universal single-mode lasing in fully chaotic 2D microcavity lasers. 

\section{Fundamental equations}
For modeling the microcavity, we assume that it is wide in the
$xy$-directions and thin in the $z$-direction.
This allows us to separate the electromagnetic fields into
transverse-magnetic (TM) and transverse-electric (TE) modes. Here we
focus only on TM modes, whose electric field vector is expressed as
$\EE$ $=$ $(0,0,E_z)$.
We also assume that the atoms in the lasing medium have spherical
symmetry and two energy levels.
The relaxation due to the interaction with the reservoir 
can be described phenomenologically 
with decay constants $\gamma_\perp$ for the microscopic polarization $\rho$   
and $\gamma_{\parallel}$ for the population inversion $W$. 
We also need to include phenomenologically the effect of the external energy 
injected into the lasing medium by the pumping power $W_{\infty}$. 

By applying the slowly varying envelope approximation, one can reduce
the Maxwell equation as follows,
\begin{equation}
 \frac{\partial}{\partial t} \tilde{E} =
 \frac{i}{2} \left( \nabla^2_{xy} + 1 \right)
 \tilde{E}
-\alpha_L \tilde{E}
 +\frac{2 \pi N \kappa \hbar}{\varepsilon} \tilde{\rho}  ,
 \label{SB-in}
\end{equation}
where $\tE$ and $\trho$ are respectively the slowly varying envelopes
of the $z$-component of the electric field and the microscopic
polarization, $N$ is the number density of the atoms, $\kappa$ is the
coupling strength, $\varepsilon$ is the permittivity, and $\alpha_L$
represents the losses describing absorption inside the cavity.  In the
above, space and time are made dimensionless by the scale
transformation $((n_{in}\omega_s/c)x,$
$(n_{in}\omega_s/c)y)$$\rightarrow$ $(x,y)$, $t\omega_s \rightarrow$
$t$, respectively, where $n_{in}$ denotes the effective refractive
index inside the cavity and $\omega_s$ is the oscillation frequency of
the fast oscillation part of the electric field.  In the same way, we
have the equation for the electric field outside the cavity,
\begin{equation}
\frac{n^2_{out}}{n_{in}^2} \frac{\partial}{\partial t} \tilde{E} =
 \frac{i}{2} \left(\nabla^2_{xy} + \frac{n^2_{out}}{n_{in}^2} \right)
 \tilde{E},
 \label{SB-out}
\end{equation}
where $n_{out}$ denotes the refractive index outside the cavity.
For the boundary condition at infinity, we adopt the outgoing wave
condition.

The optical Bloch equations are also transformed to the following form:  
\begin{equation}
 \frac{\partial}{\partial t} \tilde{\rho} =
 - \tilde{\gamma}_\perp \tilde{\rho}
 - i\Delta_0\tilde{\rho}
 + \tilde{\kappa} W \tilde{E},
\label{SB-Bloch1}
\end{equation}
\begin{equation}
 \frac{\partial}{\partial t} W =
 - \tilde{\gamma}_{\parallel} \left( W - W_\infty \right)
 - 2 \tilde{\kappa} \left( \tilde{E} \tilde{\rho}^*
 + \tilde{E}^* \tilde{\rho} \right),
\label{SB-Bloch2}
\end{equation}
where the dimensionless parameters are defined as follows: 
$\tilde{\gamma}_\perp \equiv \gamma_\perp / \omega_s$ ,
$\tilde{\gamma}_{\parallel} \equiv \gamma_{\parallel} / \omega_s$ , 
$\Delta_0 \equiv [\omega_0-\omega_s]/\omega_s$, and 
$\tilde{\kappa} \equiv \kappa / \omega_s$ has the dimension 
of the inverse of the electric field and $\omega_0$ is 
the transition frequency of the two-level atoms. 

A theoretical method to obtain stationary-state solutions of
Eq.~(\ref{SB-in})$\sim$Eq.~(\ref{SB-Bloch2}) has been developed as
``steady-state ab initio laser theory (SALT)'' \cite{Tureci2006,
  Tureci2007, HakanScience etc}.
However, the existence of a stationary-state solution does not always mean
its experimental observability.
That is, a stationary-state solution must be stable so that it can be
experimentally observed, especially when the experiment is performed
with continuous-wave operation, where the long-term dynamical effect is
expected to be important.
Such a dynamical stability is not explicitly incorporated in the SALT
approach.

In the following, we carry out the stability analysis of a
stationary-state solution for the Maxwell-Bloch equations.
Although the equations for the stability analysis are very complicated
in general, they turn out to be greatly simplified thanks to full
chaoticity and the short wavelength limit.  By applying the analysis
to a 2D microcavity laser where the ray dynamics are fully-chaotic, we
show that at least one single-mode lasing state is stable and all of
the multimode lasing states are unstable when the size of the cavity
is much larger than the wavelength and the pumping power is
sufficiently large.

\section{Dynamics of almost stationary lasing states}
We assume that near a stationary state 
the light field and polarization can be expressed as follows:   
\begin{equation}
\tE=\sum_{i}E_i (t) e^{-i  {\Delta}_i t} U_i(x,y) ,
\end{equation}
\begin{equation}
\trho=\sum_{i}\rho_i (t) e^{ -i {\Delta}_i t } V_i(x,y) ,
\end{equation}
where ${\Delta}_i$ represents the lasing oscillation frequency. 
Note that the lasing mode $i$ depends on the pumping power and 
can be a fusion of several modes that coalesce by frequency-locking,  
and separate into individual lasing modes with different frequencies 
as the pumping power decreases 
\cite{Sunada1,HarayamaAsymmetricPRL}. $U_i$ is supposed to be normalized. 

Then, from Eq.~(\ref{SB-in}), we obtain  
\begin{eqnarray}
\lefteqn{
\frac{d E_i(t)}{d t} 
+\sum_{j\neq i} \frac{d E_j(t)}{d t} e^{ -i \Delta_{ij} t }  U_{ij}  }
\nonumber \\ & & 
=  \left\{ i\left(\Delta_i +\frac{1}{2}\right)
-\left( \alpha_L +\tilde{\gamma}_{ii} \right) \right\} E_i(t)  
\nonumber \\ & & 
+\sum_{j\neq i}  
\left[
\left\{ i\left(\Delta_j +\frac{1}{2}\right) - \alpha_L \right\} U_{ij}
-\tilde{\gamma}_{ij} 
\right]
E_j(t)  e^{ -i \Delta_{ij} t } 
\nonumber \\ & &
+\frac{ 2\pi N \kappa \hbar }{\epsilon} \sum_j  e^{ -i \Delta_{ij} t }
\rho_j(t) \int_D U_i^* (x,y) V_j (x,y) dxdy , \nonumber \\ & & 
\label{E_i}
\end{eqnarray}
where $\Delta_{ij}$ denotes the frequency difference between modes
${j}$ and ${i}$, i.e., $\Delta_{ij}\equiv \Delta_j - \Delta_i$ and
$U_{ij}$ is defined by the inner product $U_{ij}\equiv\int_D U_i^*
(x,y) U_j (x,y) dxdy$, and it will be shown later that
$\tilde{\gamma}_{ii}$ is related to the flux of the light field
intensity from inside to outside the cavity through the cavity edge
and $\tilde{\gamma}_{ij}$ are defined as follows:
\begin{equation}
\tilde{\gamma}_{ij} \equiv -\frac{i}{2}\int_D dxdy 
U_i^*(x,y)\nabla^2 U_j(x,y) .
\end{equation}
In the above, $D$ denotes the area inside the cavity.

From Eq.~(\ref{SB-Bloch1}),  we have
\begin{equation}
\rho_j(t)V_j(x,y)=\frac{\tilde{\kappa}W}{\tgamma_{\perp}-i\Delta_{0j}}
E_j(t) U_j(x,y) .
\label{rho_i}
\end{equation}

\noindent 
Therefore, when the light field is almost stationary, from
Eqs.~(\ref{SB-Bloch2}) and (\ref{rho_i}), one can express the
population inversion $W$ by the light field amplitudes $E_i$ and the
spatial patterns $U_i$, i.e.,
\begin{eqnarray}
\lefteqn{W=W_\infty / } \nonumber  \\ & & 
\left[
1+
\left\{
\sum_i \sum_j 
\frac{
2\tilde{\kappa}^2 E_i E_j^* U_i U_j^* e^{-i\Delta_{ji} t }
}
{
\left\{
\tgamma_{\perp}+i\Delta_{0j} 
\right\} 
\left\{ \tgamma_{\parallel}-i\Delta_{ji} \right\}
}
 + c. c.
\right\}
\right]. \nonumber \\ 
\label{stationary W}
\end{eqnarray}
The conventional approach to treat the nonlinear terms in $W$ is to
perturbatively expand the right-hand side of Eq.~(\ref{stationary W})
for small light field amplitudes \cite{L2, Lamb}.  This method is only
applicable to just above the lasing threshold and cannot correctly
describe the case when the external pumping power $W_\infty$ is very
large.  In the following, we present a different approach applicable
to the high-pumping cases.  Note that our method can be applied to
semiconductor lasers in the same way as the conventional approach
\cite{Sargent}.

We introduce the dimensionless quantities $L$ and $C$ related to the
total intensity and mode interference, respectively, as follows:
\begin{equation}
L(x,y)\equiv
1+\sum_m a_m \left| U_m \right|^2,  
\label{I}
\end{equation} 
where $a_m$ denotes the dimensionless light field intensity of the mode $m$ weighted by 
the Lorentzian gain 
$g(\Delta_m)\equiv  \tgamma_{\perp} / (\tgamma_{\perp} ^2 + \Delta_{0m}^2)$, i.e., 
$a_m  \equiv 
({4\tilde{\kappa}^2 }/{\tgamma_{\parallel}}) g(\Delta_m) \left| E_m \right|^2$, and
\begin{equation}
C(x,y)\equiv 
\sum_{\substack{l,j \\ l\neq j} }
\frac{
2\tilde{\kappa}^2 E_l E_j^* U_l U_j^* e^{-i\Delta_{jl} t }
}
{
\left\{
\tgamma_{\perp}+i\Delta_{0j} 
\right\} 
\left\{ \tgamma_{\parallel}-i\Delta_{jl} \right\}
} 
+ c. c. 
\end{equation}
Then the denominator of the term in the right-hand side in
Eq.~(\ref{stationary W}) is expressed as $L+C=L(1+C/L)$.  The basic
idea of our approach is to expand it in the power series of $C/L$
under the condition $|C|/|L|<1$ almost everywhere in the cavity.  From
Eqs.~(\ref{E_i}), (\ref{rho_i}) and (\ref{stationary W}), we obtain
\begin{multline}
\frac{d E_i(t)}{d t} 
+\sum_{j\neq i} \frac{d E_j(t)}{d t} e^{ -i \Delta_{ij} t } U_{ij} 
\\  
=  \left\{ i\left(\Delta_i +\frac{1}{2}\right)
-\left( \alpha_L +\tilde{\gamma}_{ii} \right) \right\} E_i(t)  
\\
+\sum_{j\neq i}  
\left[
\left\{ i\left(\Delta_j +\frac{1}{2}\right) - \alpha_L \right\} U_{ij}
-\tilde{\gamma}_{ij} 
\right]
E_j(t)  e^{ -i \Delta_{ij} t }  
\\ 
+\xi W_\infty
\sum_k 
\frac{
 e^{ -i\Delta_{ik} t }
E_k 
}
{
\tgamma_{\perp}-i\Delta_{0k} 
} 
\\  
\times \int_D dx dy  
\frac{ U_i^* U_k }{ L(x,y) } 
\left[
1-\frac{C(x,y)}{L(x,y)}+\left\{\frac{C(x,y)}{L(x,y)}\right\}^2
-\cdots
\right],
\label{C/I expansion}
\end{multline}
where 
$
\xi\equiv 2\pi N \kappa \tilde{\kappa} \hbar / \varepsilon 
$.
Since we are focusing on the vicinity of the stationary state of the
slowly varying envelope, $dE_i(t)/dt$ is very small.  Therefore we can
assume $E_i(t)\sim e^{\epsilon_i t}$ where $\epsilon_i \ll
|\Delta_{ij}| $ for all $i$ and $j$ ($j\neq i$).  When Eq.~(\ref{C/I
  expansion}) is integrated over $t$, the second terms on both sides
have the coefficients of $1/\left({\epsilon_j -i\Delta_{ij}}\right)$
while the first terms $1/{\epsilon_i}$. Consequently, the
contributions of the terms concerning fast oscillations like the
second terms are much smaller than those of the first terms.
Accordingly, one can ignore the terms oscillating faster than
$e^{-i\Delta_{ik} t }$.

By ignoring the terms oscillating faster than $e^{-i\Delta_{ik} t }$,
Eq.~(\ref{C/I expansion}) is reduced to
\begin{multline}
\frac{d E_i}{d t}  \simeq  
\left\{ i\left(\Delta_i +\frac{1}{2}\right)
-\left( \alpha_L +\tilde{\gamma}_{ii} \right) \right\} E_i \\ 
+ 
\frac{ 
\xi   W_\infty  E_i 
}
{
\tgamma_{\perp}-i\Delta_{0i} 
} 
\int_D dx dy  
\frac {\left|U_i\right|^2} { L(x,y) }
\\ 
- 
\xi   W_\infty  E_i  
\int_D dxdy
\frac{ \left|U_i \right|^2 }
{
\left\{ L(x,y)\right\}^{2}
}
\sum_{\substack{k \\ k\neq i} }
\frac{ 2\tilde{\kappa}^2  \left|E_k\right|^2 \left|U_k\right|^2 }
{
(\tgamma_{\perp}-i\Delta_{0k})( \tgamma_{\parallel}-i\Delta_{ki} )
} 
\\ 
\times 
\left(
\frac{
1
}
{
\tgamma_{\perp}+i\Delta_{0k}
} 
+
\frac{
1
}
{
\tgamma_{\perp}-i\Delta_{0i}
} 
\right)
.
\end{multline}
Therefore, we obtain 
\begin{multline}
\frac{d |E_i|^2}{dt} = \frac{d}{dt} (E_i Ei^*) = E_i^* \frac{dE_i}{dt} + E_i \frac{dE_i^*}{dt} \\ 
= (-2\alpha_L + \tilde{\gamma}_{ii}+\tilde{\gamma}_{ii}^* ) |E_i|^2 
+2\xi W_{\infty} g(\Delta_i) \int_D dxdy \{L(x,y)\}^{-2} \\ 
\times |E_i|^2 |U_i|^2 
\Biggl[ L(x,y) -\sum_{k, k\neq i} \frac{2\tilde{\kappa}^2}{\tilde{\gamma}_{\perp} } 
g(\Delta_k) g_{\parallel}(\Delta_i - \Delta_k)  \\ 
\times \Bigl\{ 2\tilde{\gamma}_{\perp}    
+ (\Delta_i-\Delta_0)(\Delta_i-\Delta_k)/\tilde{\gamma}_{\perp} 
~~~~~~~~~~~~~~~~~~\\ 
+(\Delta_i-\Delta_k)(\Delta_i+\Delta_k-2\Delta_0)/\tilde{\gamma}_{\parallel}
\Bigr\}  
|E_k|^2 |U_k|^2 \Biggr],
\label{raw eq.}
\end{multline}
where $g_{\parallel}(\Delta_i - \Delta_k)$ is a Lorentzian defined as 
$g_{\parallel}(\Delta_i - \Delta_k) 
\equiv \tilde{\gamma}_{\parallel} / 
\{ \tilde{\gamma}_{\parallel}^2+(\Delta_i - \Delta_k)^2 \}$.
If $\Delta_i$ is far from $\Delta_0$, the second term of the second line 
does not contribute because $g(\Delta_i)$ almost vanishes. 
Therefore, we assume $|\Delta_i - \Delta_0|   \ll   \tilde{\gamma}_{\parallel}$.
Since the terms concerning $\Delta_k$ contribute in $L(x,y)$ 
if $\Delta_k$ is as close to $\Delta_0$ as $\Delta_i$ due to the Lorentzian $g(\Delta_k)$, 
we obtain  
\begin{equation}
L(x,y) \simeq 1+ \sum_{k, |\Delta_k - \Delta_i|  \ll \tilde{\gamma}_{\parallel} } 
\frac{4\tilde{\kappa}^2 }{\tilde{\gamma}_{\parallel}} g(\Delta_k)
|E_k|^2 |U_k|^2.
\end{equation}
Because of the Lorentzian $g_{\parallel}(\Delta_i - \Delta_k)$, 
only the terms whose $\Delta_k$ values are close to $\Delta_i$ such that 
$|\Delta_i - \Delta_k|\ll \tilde{\gamma}_{\parallel}$ contribute to  
the sum over $k$ in Eq.~(\ref{raw eq.}).  
Accordingly, we have 
$g_{\parallel}(\Delta_i - \Delta_k) \simeq 1/{\tilde{\gamma}_{\parallel}}$, 
and 
\begin{multline} 
2\tilde{\gamma}_{\perp} \gg     
(\Delta_i-\Delta_0)(\Delta_i-\Delta_k)/\tilde{\gamma}_{\perp} \\ 
+(\Delta_i-\Delta_k)(\Delta_i+\Delta_k-2\Delta_0)/\tilde{\gamma}_{\parallel}. 
\end{multline}
Consequently, we obtain  
\begin{multline}
\frac{d |E_i|^2}{dt}  
= (-2\alpha_L + \tilde{\gamma}_{ii}+\tilde{\gamma}_{ii}^* ) |E_i|^2 \\ 
+2\xi W_{\infty} g(\Delta_i) \int_D 
dxdy \{L(x,y)\}^{-2}  |E_i|^2 |U_i|^2 \\ 
\times 
\Biggl\{ 1+ 
\sum_{k, |\Delta_k - \Delta_i|  \ll \tilde{\gamma}_{\parallel} } 
\frac{4\tilde{\kappa}^2 }{\tilde{\gamma}_{\parallel}} g(\Delta_k) 
|E_k|^2 |U_k|^2 \\ 
-\sum_{k, k\neq i, |\Delta_k - \Delta_i|  \ll \tilde{\gamma}_{\parallel} } 
\frac{4\tilde{\kappa}^2 }{\tilde{\gamma}_{\parallel}} g(\Delta_k) 
|E_k|^2 |U_k|^2 \Biggr\} \\ 
= (-2\alpha_L + \tilde{\gamma}_{ii}+\tilde{\gamma}_{ii}^* ) |E_i|^2~~~~~~~~~~~~~~~~~~~~ \\ 
+2\xi W_{\infty} g(\Delta_i) \int_D 
dxdy \{L(x,y)\}^{-2}  |E_i|^2 |U_i|^2 \\ 
\times 
\Biggl\{ 1+  
\frac{4\tilde{\kappa}^2 }{\tilde{\gamma}_{\parallel}} g(\Delta_i) 
|E_i|^2 |U_i|^2
\Biggr\}. ~~~~~~~~~~~~~
\end{multline}
Therefore, we finally obtain the equation for the time evolution of
the light field intensity $I_i \equiv |E_i|^2$ of the lasing mode $i$,
\begin{equation}
\frac{d I_i }{d t}  \simeq S_i  I_i , 
\label{main1}
\end{equation}
where $S_i$ denotes the balance of the loss, gain and saturation of the mode $i$, 
and is defined as 
\begin{equation}
S_i  \equiv   -2\left( \alpha_L +\gamma_i \right) 
+2\xi W_\infty g(\Delta_i)
\int_D dx dy  
\frac{
\left|U_i \right|^2   L_i(x,y)
}
{
\left\{ L(x,y)\right\}^{2} 
},
\label{S_i}
\end{equation}
and  
$L_i(x,y)$ is related to the dimensionless light field intensity of the mode $i$, i.e.,  
$
L_i(x,y)\equiv  1+a_i \left| U_i \right|^2.
\label{I_i}
$
$\gamma_i$ is derived by applying Green's theorem to 
$(\tilde{\gamma}_{ii}+\tilde{\gamma}_{ii}^*)$  and represents the rate of the flux 
of the light field intensity going outside the cavity 
through the cavity edge for the lasing mode $i$:
\begin{equation}
\gamma_i \equiv -\frac{i}{4} \oint_{\partial D} ds \left( 
U_i^*\frac{\partial U_i}{\partial n}-U_i\frac{\partial U_i^*}{\partial n}  \right),
\end{equation}
where $\partial/\partial n$ is a normal derivative on the cavity edge.
It is important to note that Eq.~(\ref{main1}) is derived without any
assumption of small intensities, and hence it can be applied to the
strongly pumped regimes.

\section{Stability analysis of stationary lasing states}
\subsection{Stability matrix}
\label{sect:resonant_modes}
The light field intensities that make the right-hand side of Eq.~(\ref{main1}) vanish 
correspond to stationary-state solutions. 
The stability of a stationary-state solution is  
evaluated by the time evolution of the small displacements $\delta I_i$  
from the intensities $I_{s,i}$ for the stationary-state 
subject to the differential equations $d \mbox {\boldmath $\delta I$}/ dt = 
\tilde{M} \mbox {\boldmath $\delta I$}$. Here the displacement vector is defined by   
$\mbox {\boldmath $\delta I$ } \equiv {}^t 
(\delta I_1~ \delta I_2~ \cdots~ \delta I_n)$
and the matrix $\tilde{M}$ is given by 
$
\tilde{M}\equiv P M P^{-1}, 
$
and 
\begin{multline}
{M}_{ij} \equiv  
\left[ S_j +2 \xi W_\infty b_j^2 \int_D dx dy  
\frac{\left| U_j \right|^4}{\left\{ L(x,y)\right\}^{2}} \right] \delta_{ij} \\ 
-4 \xi W_\infty b_i b_j
\int_D dx dy  
\frac{ \left| U_i \right|^2 \left| U_j \right|^2 L_i(x,y) }{ \left\{L(x,y)\right\}^{3}}, 
\label{M}
\end{multline}
where 
$P\equiv \mbox{diag}(|E_1|,\cdots,|E_N|)$
and 
$
b_i \equiv (a_i g(\Delta_i))^{1/2}= 
2 \tilde{\kappa} g(\Delta_i)|E_i|/ \tgamma_{\parallel}^{1/2}
$.
For simplicity,  
we introduced the matrix $P$ and assumed its inverse matrix $P^{-1}$ exists. 
However, $|E_i|^{-1}$ in $P^{-1}$ always cancels out $|E_i|$
in $P$. Therefore, $\tilde{M}$ is always well-defined 
and the following discussion is valid irrespective of the existence of $P^{-1}$.

\subsection{Single-mode lasing states}
From Eq.~(\ref{main1}), one can see that 
the fixed point $(0,\cdots,0,I_{s,j},0,\cdots,0)$ which satisfies 
$S_j=0$ 
\noindent 
corresponds to the single-mode lasing state of the mode $j$ whose light field intensity  
is equal to $I_{s,j}=|E_{s,j}|^2$. 
Then the matrix $\tilde{M}$ for this fixed point becomes a diagonal matrix, i.e.,  
\begin{equation}
\tilde{M}_{jj}=-2 \xi W_\infty b_j^2 \int_D dx dy  
\frac{\left| U_j \right|^4}{L_{s,j}^{2}} <0, 
\label{single-diagonal}
\end{equation}
and for $i\neq j$,
\begin{equation}
\tilde{M}_{ii}= 2\xi W_\infty g(\Delta_i)  
\int_D dx dy    
\left|U_i \right|^2 
\frac{ L_{s,i}-L_{s,j}^2}{L_{s,j}^2  L_{s,i}},
\label{single-diagonal-2}
\end{equation}
where $L_{s,i}$ is related to the light intensity of the single-mode lasing 
corresponding to the fixed point $(0,\cdots,0,I_{s,i},0,\cdots,0)$, i.e.,   
$L_{s,i}\equiv 1+a_{s,i} \left| U_i \right|^2$ and 
$a_{s,i}\equiv 
({4\tilde{\kappa}^2 }/{\tgamma_{\parallel}}) g(\Delta_i) I_{s,i} $. 
From Eq.~(\ref{single-diagonal}), one can see that the single-mode lasing state 
of the mode $j$ is stable in the direction ${}^t (0~\cdots~0~\delta I_{j}~0~\cdots~0)$.

It is important to note that $L_{s,i(j)}$ contains 
the spatial pattern of the mode $i(j)$. 
As shown in the section \ref{single}, 
if the spatial pattern $|U_i|^2$ overlaps with $|U_j|^2$ inside the cavity, 
$\tilde{M}_{ii}$ can be negative.  
Then, the single-mode lasing state of the mode $j$ can be stable in the direction 
${}^t (0~\cdots~0~\delta I_{i}~0~\cdots~0)$.

\subsection{Multimode lasing states}

From Eq.~(\ref{main1}), one can see that a multimode lasing state 
corresponds to the solutions $I_{s,i}$ 
of the simultaneous equations $S_i=0$  $(i=1,2,\cdots,N)$. 
The number $N$ of the lasing modes that have nonzero light field intensities 
is an arbitrary natural number more than 1 
because $I_i=0$ always satisfies the stationary-state condition for Eq.~(\ref{main1}). 
Then, from Eq.~(\ref{M}), we have   
\begin{equation}
{M}_{ii} =  
2 \xi W_\infty b_i^2 \int_D dx dy  
\frac{\left| U_i \right|^4 L_i(x,y)}{\left\{ L(x,y) \right\}^3}
\left\{ \frac{L(x,y)}{L_i(x,y)}-2 \right\},
\label{multi-diagonal}
\end{equation}
\begin{equation}
{M}_{ij} =  
-4 \xi W_\infty b_i b_j
\int_D dx dy  
\frac{ \left| U_i \right|^2 \left| U_j \right|^2 L_i(x,y) }{ \left\{L(x,y)\right\}^{3}}. 
\label{multi-off-diagonal}
\end{equation}
\noindent 
Consequently, if and only if all of the eigenvalues of the $N\times N$ matrix 
$
\tilde{M}(\equiv P M P^{-1}) 
$
are negative, this multimode lasing state is stable. 

\section{Fully chaotic 2D microcavity lasers}
\subsection{Spatial patterns of stationary lasing modes}\label{Spatial pattern}
For bounded chaotic systems, theories on quantum ergodicity have shown 
that the probability density of finding a quantum particle in a small area 
whose state is described by the eigenfunction of 
a quantized fully chaotic system approaches 
a uniform measure as the energy of the particle increases 
and the wavelength becomes shorter 
\cite{Shnirelman etc}. 

For open chaotic mapping systems, it has been shown that the
long-lived eigenstates tend to be localized on the forward trapped set
of the corresponding classical dynamics as the wavelength decreases
\cite{Keating etc}.
This tendency can be considered as a manifestation of quantum
ergodicity in open systems.

A similar tendency has also been numerically observed for chaotic 2D
microcavities \cite{Sunada2}, where resonance wave functions of low-loss modes are
supported by the forward trapped set  
of the corresponding ray dynamics with Fresnel's law 
\cite{HaraShinoPRE, Altmann} 
in the short wavelength limit.
Because of this property, the overlap of wave functions between an
arbitrary pair of low-loss modes takes a large value
\cite{Sunada2}.
Therefore, we assume that the spatial patterns of the wave functions
for low-loss modes are similar to each other and expressed as
$
|U_i(x,y)|^2=\left\{ 1+\epsilon_i (x,y) \right\} |U_0(x,y)|^2
$, 
which implies 
\begin{equation}
\int_D dxdy \epsilon_i(x,y) |U_0(x,y)|^2=0, 
\label{overlap}
\end{equation}
because of the normalization for all of the spatial patterns of the lasing modes. 
We assume that the fluctuation $\epsilon_i (x,y)$ is so small almost everywhere 
and random that 
$
|\epsilon_i(x,y)|U_0(x,y)|^2 | \ll 1
$ and 
$
\int_D dxdy \epsilon_i(x,y) \simeq 0
$.
We also assume 
$
\int_D dxdy \epsilon_i(x,y) |U_0(x,y)|^{-2} \simeq 0
$.

\subsection{Stability of single-mode lasing states}\label{single}
In the case of a fully chaotic 2D microcavity, one can apply the wave function property 
in (\ref{overlap}) to evaluate $\tilde{M}_{ii}$ in Eq.~(\ref{single-diagonal-2})  
whose sign determines the stability of the single-mode lasing state of the mode $j$ 
in the direction ${}^t (0~\cdots~0~\delta I_{i}~0~\cdots~0)$. 
Indeed, $\tilde{M}_{ii}$ is reduced 
for the first order of 
$\epsilon_i(x,y) |U_0(x,y)|^2$
as follows:    
\begin{multline}
\tilde{M}_{ii}
= 2\xi W_\infty g(\Delta_i)
\int_S dxdy (1+\epsilon_i)|U_0|^2 \\ 
\times 
\frac{ a_{s,i} (1+\epsilon_i) -a_{s,j}^2(1+\epsilon_j)^2 |U_0|^2 }
{ a_{s,j}^2(1+\epsilon_j)^2  (1+\epsilon_i) a_{s,i}  |U_0|^6 }
|U_0|^2 \\ 
=
\frac{2\xi W_\infty g(\Delta_i)}
{ a_{s,j}^2 a_{s,i} }
\int_S dx dy    \Bigl[ 
\left\{ 1+O(\epsilon_i (x,y) |U_0(x,y)|^2) \right\}\frac{a_{s,i}}{|U_0|^2} \Bigr. \\ \Bigl. 
-\left\{ 1+O(\epsilon_j (x,y) |U_0(x,y)|^2) \right\}a_{s,j}^2 \Bigr] \\ 
\simeq
\frac{2\xi W_\infty g(\Delta_i)}{ a_{s,j}^2 a_{s,i} } 
\left[ a_{s,i} \int_S dx dy \frac{1}{|U_0|^2} -A a_{s,j}^2 \right] 
,
\label{single-diagonal-final}
\end{multline}
where $S$ denotes the support of $|U_0(x,y)|^2$ and $A$ is its area. 

If and only if $\tilde{M}_{ii}$ is negative, the single-mode lasing state of the mode $j$ 
is stable in the direction ${}^t (0~\cdots~0~\delta I_{i}~0~\cdots~0)$. 
Therefore, from  Eq.~(\ref{single-diagonal-final}),
we obtain the stability condition for the single mode lasing of the mode $j$, 
\begin{equation}
\left( \frac{a_{s,j}}{A} \right)^2 > 
\frac{a_{s,i}}{A} \int_S dxdy \frac{1}{ A^2 |U_0|^2} .    
\label{single mode stability condition}
\end{equation} 
\noindent
Accordingly, the single-mode lasing state of the mode $j$ is stable 
when its intensity is large enough to satisfy 
Eq.~(\ref{single mode stability condition}) for all of the other modes $i$ values.  
Note that the integral in Eq.~(\ref{single mode stability condition}) 
is estimated to be approximately unity 
because $|U_0|^{-2}$ can be approximated to be $A$. 
Consequently,  the mode that has the largest single-mode intensity 
is stable, and hence there always exists one stable single-mode lasing state at least. 

\subsection{Instability of multimode lasing states}
Next, we show that all of the multimode lasing states are unstable
under the assumptions that all of the spatial patterns $U_i$ of 
the single-mode lasing states are similar to each other,
and the pumping power $W_\infty$ is very high,   
which implies that the light field intensities are very large. 

According to the assumption in Section \ref{Spatial pattern} 
for the spatial patterns of the lasing modes 
in a fully chaotic microcavity, we obtain 
\begin{eqnarray}
L_i(x,y)  
& \simeq & (1+\epsilon_i) a_i |U_0|^2, 
\end{eqnarray}
\begin{eqnarray}
L(x,y)
& \simeq & 
\sum_{m=1}^N(1+\epsilon_m) a_m |U_0|^2  
\end{eqnarray}
for the support $S$ of $|U_0|^2$ and we assumed $a_i\gg 1$.
Then, the matrix elements in Eqs.~(\ref{multi-diagonal}) and (\ref{multi-off-diagonal})
can be expressed by using  the fluctuations $\epsilon_i(x,y) $.
Therefore, from the properties assumed for $\epsilon_i(x,y) $, 
we obtain for the first order of $\epsilon_i(x,y) |U_0(x,y)|^2$, 
\begin{equation}
{M}_{ii} \simeq  
-4 \xi A W_\infty b_i^2   
\frac{a_i}{a_{tot}^3}
\left( 1-\frac{a_{tot}}{2a_i} \right),
\label{multi-diagonal2}
\end{equation}
\begin{equation}
{M}_{ij} \simeq  
-4 \xi A W_\infty b_i b_j
\frac{a_i}{a_{tot}^3}, 
\label{multi-off-diagonal2}
\end{equation}
where $a_{tot}\equiv \sum_{m=1}^N a_m$.  

It is important to note that 
$\tilde{M}(N)$ can be factorized as 
$
\tilde{M} (N) =  
-4\xi A W_{\infty}/( a_{tot}^3 )
P Q B R B P^{-1},
$
where 
$
Q \equiv \mbox{diag}(a_1,\cdots,a_N),
$
$
B \equiv \mbox{diag}(b_1,\cdots,b_N),
$
and the diagonal elements of $R$ are given by 
$
R_{ii} \equiv 1-{a_{tot}}/{(2a_i)} ,
$
while the off-diagonal elements
$
R_{ij} =1 
$
.
Then, as is explained in Appendix, all of the eigenvalues of $\tilde{M}(N)$
are equal to those of $M^{\prime}(N)$ defined as 

\begin{multline}
M^\prime(N) \equiv  
-\frac{
4\xi A W_{\infty}  
}
{ a_{tot}^3 }
\mbox{ diag } (\sqrt{a_1},\cdots,\sqrt{a_N}) \\ \times 
B R B \mbox{ diag } (\sqrt{a_1},\cdots,\sqrt{a_N}).
\end{multline}

\noindent
All of the eigenvalues of $M^{\prime}(N)$ are real because it is a real symmetric matrix.  
According to the theorem of linear algebra, 
the number of the negative eigenvalues of $M^{\prime}(N)$ is equal to that of 
sign changes of the sequence 
$\{1,|M^{\prime}(1)|,|M^{\prime}(2)|,\cdots,|M^{\prime}(N)| \}$ 
where the minor determinants of $M^{\prime}(N)$ are given as explained in Appendix, 
\begin{equation}
\left|M^\prime(k)\right|
=  
\left(
\frac{
2\xi A W_{\infty}
}
{ a_{tot}^2 }
\right)^k
\left( 1-\sum_{i=1}^k \frac{2a_i}{a_{tot}} \right)
\prod_{i=1}^k b_i^2 .
\label{M'-determinant}
\end{equation}
\noindent 
Since the term  $(1-\sum_{i=1}^k {2a_i}/{a_{tot}})$ 
decreases monotonically for $k$ and equals $-1$ when $k=N$, 
the above sequence of the minor determinants of $M^{\prime}(N)$ 
changes the sign once.  Accordingly, $\tilde{M}$ has one negative 
eigenvalue and $(N-1)$ positive eigenvalues, which means the fixed point 
corresponding to the multimode lasing state is an unstable saddle point.

\section{Summary and discussion}
By introducing an expansion method 
different from the conventional theories \cite{Lamb, L2, Sargent} 
for the population inversion in the Maxwell-Bloch equations and evaluating the
eigenvalues of a stability matrix describing the interactions
among a huge number of lasing modes, we theoretically showed
that in a fully chaotic 2D microcavity laser, at least one
single-mode lasing state is stable, while all multimode lasing states
are unstable, when the external pumping power is sufficiently large and 
the cavity size is much larger than the wavelength 
to the extent that $ \tgamma_{\parallel}\gg |\Delta_{ij}| $, where $\Delta_{ij} $ 
is the difference between the adjacent lasing frequencies.
This result provides a theoretical ground for recent experimental
observations of universal single-mode lasing in fully chaotic 2D
microcavity lasers \cite{Sunada1, Sunada2}.

It is important to note that the theory presented in this paper should 
be applied to explain the observation of single-mode lasing in 
the experiments of continuous-wave pumping cases. 
Generally, the lifetime of an unstable multimode lasing state can be much longer
than a pulse width. Thus, for a pulsed operation, it is likely that
the collapse of an unstable multimode lasing state cannot be achieved 
within a pulse width, 
even if the size of a fully chaotic 2D microcavity is sufficiently large, and  
multimode lasing is observed and universal single-mode lasing 
seems to disappear 
\cite{ReddingCao, ChoiMulti}. 
In this case, as the pulse width is increased, the number of lasing modes 
decreases \cite{Sunada1}.
Multimode lasing in a fully chaotic 2D microcavity can also be observed  
when the condition $ \tgamma_{\parallel}\gg |\Delta_{ij}| $ is not satisfied 
\cite{CerjanOPEX, Sunada3}.
This condition for multimode lasing coincides with that derived for one-dimensional 
lasers \cite{FuHaken}.
 
The theory presented in this paper cannot give the threshold pumping power 
for single-mode lasing, but it is useful for understanding the single-mode lasing 
mechanism. 
In addition, it does not take into consideration these phenomena that might 
affect lasing characteristics such as the thermal effect. 
However, according to previous studies \cite{Sunada1, Sunada2}, 
we can at least say that the threshold for single-mode lasing is achievable 
in real experiments. 
It is of interest to further demonstrate the predicted single-mode
lasing experimentally for various fully chaotic cavities. It is also
important to elucidate experimentally, numerically and theoretically
how multimode lasing states in a low pumping regime and/or in a small
cavity change into a single-mode lasing state as the pumping power
and/or the size of the cavity are increased.

\renewcommand{\theequation}{A.\arabic{equation}}
\setcounter{equation}{0}
\section*{Appendix}
Let us suppose that $\mbox {\boldmath $x$ } $ is the eigenvector 
corresponding to the eigenvalue $\lambda$ of the matrix 
$M \equiv \mbox{diag}(B_1,\cdots,B_N)C\mbox{diag}(A_1,\cdots,A_N)$ 
where the diagonal element of the matrix C is defined as $C_{ii}\equiv C_i$ 
and every off-diagonal element is equal to 1, that is, 
\begin{equation}
M \mbox {\boldmath $x$ } =
\mbox{diag}(B_1,\cdots,B_N)C\mbox{diag}(A_1,\cdots,A_N)\mbox {\boldmath $x$ } 
=\lambda \mbox {\boldmath $x$ }.
\label{appendix1}
\end{equation} 
The left-hand side of Eq.~(\ref{appendix1}) is rewritten as follows: 
\begin{multline}
\mbox{diag}(B_1,\cdots,B_N)C 
\mbox{diag}(A_1^{1/2},\cdots,A_N^{1/2})  \\ 
\times 
\mbox{diag}(A_1^{1/2},\cdots,A_N^{1/2})  
\mbox{diag}(B_1^{1/2},\cdots,B_N^{1/2}) \\   
\times \mbox{diag}(B_1^{-1/2},\cdots,B_N^{-1/2}) 
\mbox {\boldmath $x$ }~~~~~~~~~~~~~~~~~~\\ 
=\mbox{diag}(B_1,\cdots,B_N)C  
\mbox{diag}(B_1^{1/2},\cdots,B_N^{1/2})~~~~~~~~~~~~~~~~~~~~~~~ \\   
\times \mbox{diag}(A_1^{1/2},\cdots,A_N^{1/2}) 
\mbox{diag}(B_1^{-1/2},\cdots,B_N^{-1/2})  \\ 
\times \mbox{diag}(A_1^{1/2},\cdots,A_N^{1/2})  
\mbox {\boldmath $x$}.~~~~~~~~~~~~~~~~~~~~~~
\end{multline} 
On the other hand, the right-hand side of Eq.~(\ref{appendix1}) is rewritten as follows:
\begin{multline}
\lambda    
\mbox{diag}(A_1^{-1/2},\cdots,A_N^{-1/2}) 
\mbox{diag}(A_1^{1/2},\cdots,A_N^{1/2}) \\
\times 
\mbox{diag}(B_1^{1/2},\cdots,B_N^{1/2})
\mbox{diag}(B_1^{-1/2},\cdots,B_N^{-1/2})
\mbox {\boldmath $x$ } \\ 
= 
\lambda    
\mbox{diag}(A_1^{-1/2},\cdots,A_N^{-1/2}) 
\mbox{diag}(B_1^{1/2},\cdots,B_N^{1/2}) \\ 
\times \mbox{diag}(B_1^{-1/2},\cdots,B_N^{-1/2})
\mbox{diag}(A_1^{1/2},\cdots,A_N^{1/2}) \mbox {\boldmath $x$}. 
\end{multline}
Therefore, we have 
\begin{multline}
\mbox{diag}(B_1,\cdots,B_N)C  
\mbox{diag}(B_1^{1/2},\cdots,B_N^{1/2})~~~~~~~~~~~~~~~~~~~~~~~ \\   
\times \mbox{diag}(A_1^{1/2},\cdots,A_N^{1/2}) 
\mbox {\boldmath $x^{\prime}$ } \\ 
= 
\lambda    
\mbox{diag}(A_1^{-1/2},\cdots,A_N^{-1/2}) 
\mbox{diag}(B_1^{1/2},\cdots,B_N^{1/2}) 
\mbox {\boldmath $x^{\prime}$}, 
\label{appendix2}
\end{multline}
where $\mbox {\boldmath $x^{\prime}$ } \equiv 
\mbox{diag}(B_1^{-1/2},\cdots,B_N^{-1/2})
\mbox{diag}(A_1^{1/2},\cdots,A_N^{1/2}) \mbox {\boldmath $x$}$.
Operating $\mbox{diag}(B_1^{-1/2},\cdots,B_N^{-1/2})
\mbox{diag}(A_1^{1/2},\cdots,A_N^{1/2}) $ 
to both sides of Eq.~(\ref{appendix2}) yields 
\begin{multline}
\mbox{diag}(A_1^{1/2},\cdots,A_N^{1/2}) \mbox{diag}(B_1^{1/2},\cdots,B_N^{1/2})C \\  
\times \mbox{diag}(B_1^{1/2},\cdots,B_N^{1/2}) \mbox{diag}(A_1^{1/2},\cdots,A_N^{1/2}) 
\mbox {\boldmath $x^{\prime}$ }  
= 
\lambda    
\mbox {\boldmath $x^{\prime}$}.
\label{appendix3}
\end{multline}
From Eq.~(\ref{appendix3}), one can see that the eigenvalue of the matrix $M^{\prime}$ 
defined as 
\begin{multline}
M^{\prime} \equiv 
\mbox{diag}(A_1^{1/2},\cdots,A_N^{1/2}) \mbox{diag}(B_1^{1/2},\cdots,B_N^{1/2})C \\ 
\times \mbox{diag}(B_1^{1/2},\cdots,B_N^{1/2}) \mbox{diag}(A_1^{1/2},\cdots,A_N^{1/2}), 
\end{multline}
is equal to $\lambda$. 
Since $M_{ij}^{\prime}=C_{ij}(A_i A_j B_i B_j)^{1/2}$ and $C$ is a real symmetric matrix, 
$M^{\prime}$ is a real symmetric matrix. 
The determinant of $M^{\prime}$ is given as follows:
\begin{multline}
|M^{\prime}| = 
|\mbox{diag}(A_1^{1/2},\cdots,A_N^{1/2})| |\mbox{diag}(B_1^{1/2},\cdots,B_N^{1/2})||C| \\ 
\times |\mbox{diag}(B_1^{1/2},\cdots,B_N^{1/2})|
|\mbox{diag}(A_1^{1/2},\cdots,A_N^{1/2})|\\ 
=\left( \prod_{i=1}^N A_i B_i \right) \left( \prod_{i=1}^N (C_i-1) \right) 
\left( 1+\sum_{i=1}^N (C_i-1)^{-1} \right).
\end{multline}

\section*{Funding}
This work was supported in part by a Waseda University Grant for
Special Research Projects (Project number: 2017B-197).


\begin{thebibliography}{99}
%
\bibitem{StoeckmannBook}
 H.-J. St\"{o}ckmann, {\it Quantum Chaos: An Introduction} 
(Cambridge University Press, Cambridge, England, 1999).

\bibitem{Haake}
F. Haake, {\it Quantum Signaures of Chaos}  
(Springer-Verlag, New York, USA, 2000).

\bibitem{NakamuraHarayama}
K. Nakamura and T. Harayama, 
{\it Quantum Chaos and Quantum Dots} 
(Oxford University Press, Oxford, 2004).

\bibitem{BohigasCasatiBerry}
O. Bohigas, M. J. Giannoni, and C. Schmit,
``Characterization of chaotic quantum spectra and universality of level fluctuation Laws,'' 
Phys. Rev. Lett. \textbf{52}, 1-4 (1984); 
G. Casati, F. Valz-Gris, and I.Guarneri, 
``On the connection between quantization of nonintegrable systems and statistical theory of spectra,'' 
Lett. Nuovo Cimento Soc. Ital. Fis. 28, 279-282(1980); 
M.V. Berry, 
``Quantizing a classically ergodic system: Sinai's billiard and the KKR method,'' 
Ann. Phys. (N.Y.) 131, 163-216 (1981). 
%
\bibitem{MullerHeusler}
M.V. Berry, 
``Semiclassical theory of spectral rigidity,''
Proc. R. Soc. London, Ser. A {\bf 400}, 229-251(1985).; 
M. Sieber and K. Richter, 
``Correlations between periodic orbits and their role in spectral statistics,'' 
Phys. Scr. T90, 128-133 (2001);
M. Sieber, 
``'Leading off-diagonal approximation for the spectral form factor for uniformly hyperbolic systems,''
J. Phys. A 35, L613-L619 (2002);
S. M\"{u}ller, S. Heusler, P. Braun, F. Haake, and A. Altland, 
``Periodic-orbit theory of universality in quantum chaos,'' 
Phys. Rev. Lett. \textbf{93}, 014103 (2004); 
S. Heusler, S. M\"{u}ller, A. Altland, P. Braun, and F. Haake, 
``Periodic-orbit theory of level correlations,'' 
Phys. Rev. Lett. \textbf{98}, 044103 (2007).

\bibitem{Jalabert}
R. A. Jalabert, H. U. Baranger, and A. D. Stone, 
``Conductance fluctuations in the ballistic regime: A probe of quantum chaos?,'' 
Phys. Rev. Lett. \textbf{65}, 2442-2445 (1990); 
%
C. M. Marcus, A. J. Rimberg, R. M. Westervelt, P. F. Hopkins, and A. C. Gossard, 
``Conductance fluctuations and chaotic scattering in ballistic microstructures,''  
Phys. Rev. Lett. \textbf{69}, 506-509 (1992).

\bibitem{StoneNoeckeletc}
J. U. N\"ockel and A. D. Stone, 
``Ray and wave chaos in asymmetric resonant optical cavities,'' 
Nature {\bf 385},45-47 (1997); 
%
C. Gmachl,  
F. Capasso, E. E. Narimanov, J. U. N\"{o}ckel, A. D. Stone, 
J. Faist, D. L. Sivco and A. Y. Cho, 
``High-power directional emission from microlasers with chaotic resonators,'' 
Science {\bf 280}, 1556-1564 (1998);
%
S.-B. Lee, 
J.-H. Lee, J.-S. Chang, H.-J. Moon, S. W. Kim, and K. An, 
``Observation of scarred modes in asymmetrically deformed microcylinder lasers,''  
Phys. Rev. Lett. \textbf{88}, 033903 (2002);
%
S. -Y. Lee, 
S. Rim, J. W. Ryu, T. Y. Kwon, M. Choi,and C. -M. Kim,
``Quasiscarred resonances in a spiralshaped microcavity,'' 
Phys. Rev. Lett. \textbf{93}, 164102 (2004); 
%
H. G. L. Schwefel, 
N. B. Rex, H. E. Tureci, R. K. Chang, A. D. Stone, 
T. Ben-Messaoud, and J. Zyss, 
``Dramatic shape sensitivity of directional emission patterns from similarly deformed cylindrical polymer lasers,'' 
J. Opt. Soc. Am. B 21, 923-934 (2004); 
%
V. A. Podolskiy, E. Narimanov, W. Fang, and H. Cao, 
``Chaotic microlasers based on dynamical localization,'' 
Proc. Natl. Acad. Sci. USA \textbf{101}, 10498-10500 (2004); 
%
J. Wiersig, 
``Formation of Long-Lived, Scarlike Modes 
near Avoided Resonance Crossings in Optical Microcavities,'' 
Phys. Rev. Lett. \textbf{97}, 253901 (2006); 
%
J. Wiersig and M. Hentschel,
``Combining directional light output and ultralow loss in deformed microdisks,'' 
Phys. Rev. Lett. {\bf 100}, 033901 (2008); 
%
E. Bogomolny, R. Dubertrand, and C. Schmit, 
``Trace formula for dieletric cavities : I. General properties,'' 
Phys. Rev. E \textbf{78}, 056202 (2008); 
%
S. Shinohara, 
T. Harayama, T. Fukushima, M. Hentschel, T. Sasaki,
and E. E. Narimanov, 
``Chaos-assisted directional light emission from microcavity lasers,'' 
Phys. Rev. Lett. {\bf 104}, 163902 (2010); 
%
Q. Song, L. Ge, B. Redding, and H. Cao, 
``Channeling chaotic rays into waveguides for efficient collection of 
microcavity emission,'' 
Phys. Rev. Lett. {\bf 108}, 243902 (2012).
%
R. Sarma, L. Ge, J. Wiersig, and H. Cao, 
``Rotating optical microcavities with broken chiral symmetry,'' 
Phys. Rev. Lett. {\bf 114}, 053903 (2015).

\bibitem{JanCao}
H. Cao and J. Wiersig, 
``Dielectric microcavities: Model systems for wave chaos and non-Hermitian physics,'' 
Rev. Mod. Phys. {\bf 87}, 61-111 (2015).

\bibitem{HarayamaShinohara}
T. Harayama and S. Shinohara, 
``Two-dimensional microcavity lasers,''
Laser Photonics Rev. {\bf 5}, 247-271 (2011).

\bibitem{Sunada1}
S. Sunada, T. Fukushima, S. Shinohara, T. Harayama, and M. Adachi, 
``Stable single-wavelength emission from fully chaotic microcavity lasers,'' 
Phys. Rev. A {\bf 88}, 013802 (2013). 

\bibitem{Sunada2}
S. Sunada, S. Shinohara, T. Fukushima, and T. Harayama, 
``Signature of Wave Chaos in Spectral Characteristics of Microcavity Lasers,'' 
Phys. Rev. Lett. {\bf 116}, 203903 (2016).

\bibitem{Fisher}
M. Fisher, 
``The renormalization group in the theory of critical behavior,''  
Rev. Mod. Phys. {\bf 46}, 597-616 (1974).

\bibitem{Harayama2003}T. Harayama, P. Davis and K. S. Ikeda, 
``Stable oscillations of a spatially chaotic wave function in a microstadium laser,'' 
Phys. Rev. Lett. {\bf 90}, 063901 (2003).

\bibitem{Harayama2005}
T. Harayama, S. Sunada, and K. S. Ikeda, 
``Theory of two dimensional microcavity lasers,'' 
Phys. Rev. A {\bf 72}, 013803 (2005).

\bibitem{Tureci2006}
H. E. T\"{u}reci, A. D. Stone, and B. Collier, 
``Self-consistent multimode lasing theory  
for complex or random lasing media,''  
Phys. Rev. A {\bf 74}, 043822 (2006).

\bibitem{Tureci2007}
H. E. T\"{u}reci, A. D. Stone, and L. Ge, 
``Theory of the spatial structure of nonlinear lasing modes,'' 
Phys. Rev. A {\bf 76}, 013813 (2007).

\bibitem{Lamb}
W. E. Lamb, 
``Theory of an optical maser,'' 
Phys. Rev. A {\bf 134}, 1429-1450 (1964).

\bibitem{L2}
M. Sargent, III, M. O. Scully, and W. E. Lamb, Jr,
{\it Laser Physics} 
(Addison-Wesley, Reading, MA, 1974).

\bibitem{Sargent}
M. Sargent III, 
``Theory of a multimode quasiequilibrium semiconductor laser,'' 
Phys. Rev. A {\bf 48}, 717-726 (1993).

\bibitem{HarayamaAsymmetricPRL}
T. Harayama, T. Fukushima, S. Sunada and K. S. Ikeda, 
``Asymmetric stationary lasing patterns in 2D symmetric microcavities,'' 
Phys. Rev. Lett. {\bf 91}, 073903 (2003).

\bibitem{HakanScience etc}
H. E. T\"{u}reci, L. Ge, S. Rotter, and A. D. Stone, 
``Strong interactions in multimode random lasers,'' 
Science {\bf 320}, 643-646 (2008); 
%
Li Ge, R. J. Tandy, A. D. Stone, and H. E. T\"{u}reci, 
``Quantitative verification of ab initio self-consistent laser theory,'' 
Opt. Exp. {\bf 16}, 16895-16902 (2008);
%
H. E. T\"{u}reci, A. D. Stone, L. Ge, S. Rotter, and R. J. Tandy, 
``Ab initio self-consistent laser theory and random lasers,'' 
Nonlinearity {\bf 22}, C1-C18 (2009); 
%
L. Ge, Y. D. Chong, and A. D. Stone, 
``Steady-state ab initio laser theory: generalizations and analytic results,'' 
Phys. Rev. A {\bf 82}, 063824 (2010);  
%
A. Cerjan and A. D. Stone, 
``Steady-state ab initio theory of lasers with injected signals,'' 
Phys. Rev. A {\bf 90}, 013840 (2014). 
%
S. Esterhazy, 
D. Liu, M. Liertzer, A. Cerjan, L. Ge, K. G. Makris, A. D. Stone, 
J. M. Melenk, S. G. Johnson, and S. Rotter
``Scalable numerical approach for the steady-state ab initio laser theory,'' 
Phys. Rev. A {\bf 90}, 023816 (2014); 
%
A. Pick, 
A. Cerjan, D. Liu, A. W. Rodriguez, A. D. Stone, Y. D. Chong, and S. G. Johnson
``Ab initio multimode linewidth theory for arbitrary inhomogeneous laser cavities,'' 
Phys. Rev. A {\bf 91}, 063806 (2015).

\bibitem{Keating etc}
H. Schomerus and J. Tworzydlo, 
``Quantum-to-classical crossover of quasibound states in open quantum systems,'' 
Phys. Rev. Lett. {\bf 93}, 154102 (2004): 
%
S. Nonnenmacher and M. Zworski,
``Fractal Weyl laws in discrete models of chaotic scattering,''  
J. Phys. A {\bf 38}, 10683-10702 (2005); 
%
J. P. Keating, M. Novaes, S. D. Prado, and M. Sieber, 
``Semiclassical structure of chaotic resonance eigenfunctions,'' 
Phys. Rev. Lett. {\bf 97}, 150406 (2006); 
%
D. L. Shepelyansky, 
``Fractal Weyl law for quantum fractal eigenstates,'' 
Phys. Rev. E {\bf 77}, 015202 (R) (2008); 
%
M. Novaes, 
``Resonances in open quantum maps,'' 
J. Phys. A: Math. Theor. {\bf 46}, 143001 (2013).
%

\bibitem{Shnirelman etc}
A. I. Shnirelman, 
``Ergodic properties of eigenfunctions,'' 
Usp. Mat. Nauk {\bf 29}, 181-182 (1974); 
Y. Colin de Verdie`re, 
``Ergodicit\'{e} et fonctions propres du laplacien,'' 
Commun. Math. Phys. {\bf 102}, 497-502 (1985); 
S. Zelditch, 
``Uniform distribution of eigenfunctions on compact hyperbolic surfaces,'' 
Duke Math. J. {\bf 55}, 919-941 (1987); 
B. Helffer, A. Martinez, and D. Robert, 
``Ergodicit\'{e} et limite semiclassique,'' 
Commun. Math. Phys. {\bf 109}, 313-326 (1987); 
S. Zelditch and M. Zworski, 
``Ergodicity of eigenfunctions for ergodic billiards,'' 
Commun. Math. Phys. {\bf 175}, 673-682 (1996); 
A. B\"{a}cker, R. Schubert, and P. Stifter, 
``Rate of quantum ergodicity in Euclidean billiards,'' 
Phys. Rev. E, {\bf 57}, 5425-5447 (1998).

\bibitem{HaraShinoPRE}
T. Harayama and S. Shinohara, 
``Ray-wave correspondence in chaotic dielectric billiards,'' 
Phys. Rev. E {\bf 92}, 042916 (2015). 

\bibitem{Altmann}
E. G. Altmann, 
``Emission from dielectric cavities in terms of invariant sets of the chaotic ray dynamics,'' 
Phys. Rev. A {\bf 79}, 013830 (2009).

\bibitem{ReddingCao}
B. Redding, A. Cerjan, X. Huang, M. L. Lee, A. D. Stone, M. A. Choma, and H. Cao, 
``Low-Spatial Coherence Electrically-Pumped Semiconductor Laser for Speckle-Free Full-Field Imaging,'' 
Proc. Natl. Acad. Sci. USA {\bf 112}, 1304-1309 (2015).

\bibitem{ChoiMulti}
M. Choi, S. Shinohara, and T. Harayama, 
``Dependence of far-field characteristics on the number of lasing modes in stadium-shaped InGaAsP microlasers,'' 
Opt. Express {\bf 16}, 17544-17559 (2008). 


\bibitem{CerjanOPEX}
A, Cerjan, B, Redding, L. Ge, S. F. Liew, 
H. Cao, and A. D. Stone,
``Controlling mode competition by tailoring the spatial pump distribution 
in a laser: a resonance-based approach,'' 
Opt. Express {\bf 24}, 26006-26015 
(2016). 

\bibitem{Sunada3}
S. Sunada, T. Harayama, and K. S. Ikeda, 
``Multimode lasing in fully chaotic cavity lasers,'' 
Phys. Rev. E  {\bf 74},  046209 (2005).

\bibitem{FuHaken}
H. Fu and H. Haken, 
``Multifrequency operation in a short-cavity standing-wave laser,'' 
Phys. Rev. A  {\bf 43},  2446-2454 (1991).

%
\end{thebibliography}
\end{document}